\documentclass[onecolumn]{aa}
\usepackage{aas_macros}
\usepackage{natbib}
\usepackage{txfonts}
\usepackage{array}
\usepackage{caption}
\usepackage{multicol}

 \usepackage[squaren,Gray,thinspace,thickqspace]{SIunits}
\usepackage[rightcaption]{sidecap}
\usepackage{graphicx,subfigure}
\sidecaptionvpos{figure}{c} 

\usepackage{hyperref}
\hypersetup{
colorlinks=true,
citecolor=blue,
linkcolor=blue,
urlcolor=black
}

\renewcommand{\thesection}{}
\renewcommand{\thesubsection}{\arabic{subsection}}

\title{Unveiling the atmospheres of giant exoplanets with an EChO-class mission.}
\titlerunning{Unveiling the atmospheres of giant exoplanets with an EChO-class mission.}
\authorrunning{Parmentier et al.}

\author{Vivien Parmentier\inst{1}, Adam P. Showman\inst{2}, Julien de Wit\inst{3}}
\institute{Universit\'e de Nice-Sophia Antipolis, Observatoire de la C\^ote d'Azur, CNRS UMR 6202, Nice, France [vivien.parmentier@oca.eu]
\and 
Department of Planetary Sciences, Lunar and Planetary Laboratory,University of Arizona, Tucson AZ, USA
\and 
Department of Earth, Atmospheric and Planetary Sciences, MIT, 77 Massachusetts Avenue, Cambridge, MA 02139, USA.
} 

\date{Accepted to the Experimental Journal of Astronomy \today}

\begin{document}
\maketitle
\noindent {   Abstract .} More than a thousand exoplanets have been discovered {  over} the last decade. Perhaps more excitingly, probing their atmospheres {  has become} possible. With current data we {  have glimpsed} the diversity of exoplanet atmospheres that will be revealed {  over the coming decade}. However, numerous questions concerning their chemical composition, thermal structure, and atmospheric dynamics remain to be answered. More observations of higher quality are needed. In the next years, the selection of a space-based mission dedicated to the spectroscopic characterization of exoplanets would revolutionize our understanding of the physics of planetary atmospheres. { Such a mission was proposed to the ESA cosmic vision program in 2014. Our paper is therefore based on the planned capabilities of the Exoplanet Characterization Observatory (EChO), but it should equally apply to any future mission with similar characteristics. With its large spectral coverage ($4-16\, \micro\meter$), high spectral resolution ($\Delta\lambda/\lambda>300$ below $5\,\micro\meter$ and $\Delta\lambda/\lambda>30$ above $5\,\micro\meter$) and $1.5\meter$ mirror, a future mission such as EChO will provide spectrally resolved transit lightcurves, secondary eclipses lightcurves, and full phase curves of numerous exoplanets with an unprecedented signal-to-noise ratio.
In this paper, we review some of today's main scientific questions about gas giant exoplanets atmospheres, for which a future mission such as EChO will bring a decisive contribution.}

\section*{Introduction}
{ 
Characterizing exoplanets atmospheres has recently become within reach. Nowadays, a significant number of atmospheric measurements have been acquired on a dozen of exoplanets. Unfortunately, none of those measurements were done with a dedicated instrument. Although {  researchers have} made the best use of available telescopes, the observations still suffer from large error bars, from possible instrumental noise~\citep{Hansen2014}, are averaged over large bins of frequency, and measurements at different wavelength are usually made at different times. The construction of a reliable spectrum is therefore a difficult task. Few unambiguous molecular detections have been claimed and most of the physical characterizations are qualitative rather than quantitative. Better data are needed. The future of exoplanet characterization should be based on high signal-to-noise, spectrally resolved observations with a large spectral coverage accessible in a single observation. 
 
A mission with those capabilities was proposed to the ESA Cosmic Vision program in 2014. With its large spectral coverage ($4-16\, \micro\meter$), high spectral resolution ($\lambda/\Delta\lambda>300$ below $5\,\micro\meter$ and $\lambda/\Delta\lambda>30$ above $5\,\micro\meter$), and $1.5\,\meter$ mirror, EChO (the Exoplanet Characterization Observatory) is an ideal instrument to characterize exoplanets atmospheres~\citep[see][for more technical details about the mission]{Tinetti2012}. Although it was not selected in 2014, it should serve as a baseline for future missions with similar goals. The following review is based on the expected capabilities of EChO but is also relevant for any future mission with similar characteristics. In the following, the term \emph{EChO} should therefore be understood as \emph{an EChO-class mission}.

We will now review why a mission such as EChO will be a decisive step toward understanding exoplanets atmospheres and atmospheric physics in general.}

\section*{On the large diversity of observable exoplanets atmospheres}

 \begin{figure}[h!]
\begin{minipage}[c]{0.48\linewidth} 
\includegraphics[width=1\linewidth, trim=25 10 50 25]{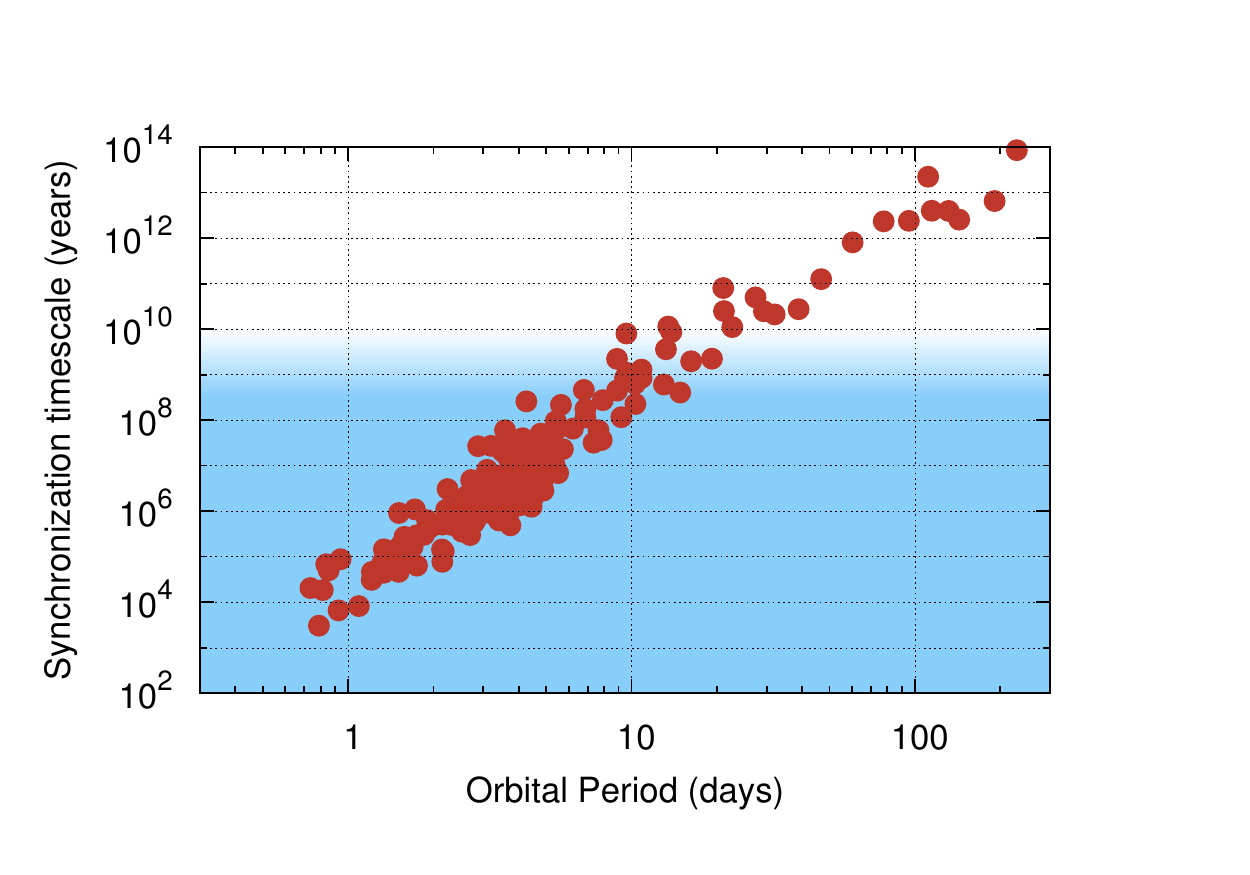}
\centering
  \caption{Tidal synchronization timescale based on~\citet{Guillot1996} for all known exoplanets with a measured mass and radius in function of their orbital period for a dissipation factor $Q=6\times10^{5}$, typical for hot Jupiters~\citep{Ferraz-Mello2013} and an initial rotation rate equal to Jupiter's one. Planets in the shaded area are likely to be tidally locked.}
\label{fig::PlotSync}
\end{minipage}
\hfill
\begin{minipage}[c]{0.48\linewidth} 
\centering
\includegraphics[width=0.85\linewidth, trim=25 10 50 25]{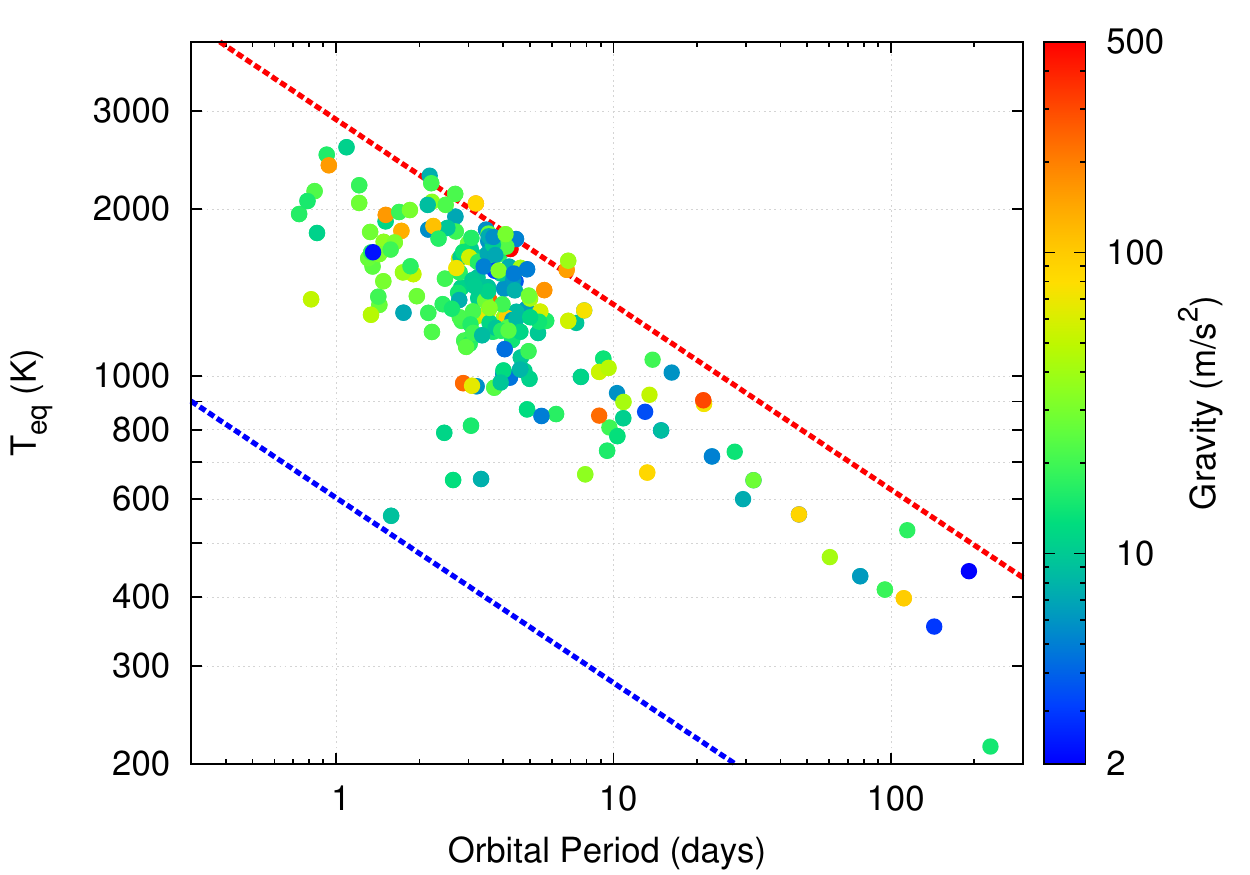}
  \caption{Equilibrium temperature (assuming zero albedo) of exoplanets with a measured mass and radius. Planets are color-coded by their gravity. The blue (red) line is the equilibrium temperature for a planet orbiting a M5 (A5) type star. Planets with an orbital period smaller than $\approx 10$ days are likely to have a rotation period equal to their orbital period (see Fig.~\ref{fig::PlotSync}).}
\label{fig::PlotTeq}
\end{minipage}
\end{figure}

Most EChO targets -- and the ones for which the best observations will be available -- are planets orbiting close to their host star. Tidal interactions should force them toward a tidally locked state~\citep{Lubow1997,Guillot2002} where their rotation period is the same as their revolution period (see Fig.~\ref{fig::PlotSync}). A whole range of atmospheric constraints is obtainable for those close-in, tidally locked planets because \emph{we know which hemisphere is facing us at any orbital phase}. Monitoring the star-planet system during its whole orbit, one can obtain longitudinal information on the planet's brightness distribution~\citep{Knutson2008}. During the ingress and egress of the secondary eclipse, the technique of eclipse mapping~\citep{Majeau2012,DeWit2012} can constrain the horizontal (both longitudinal and latitudinal) brightness distribution of the planet's dayside.  {  Finally, the frequency dependence of the thermal flux emitted by the planet and of the stellar flux filtered {  through} the planet atmosphere during transit depends principally on the temperature profile, the atmospheric composition and their variations with depth~\citep[]{Barstow2013,deWit2013}.Thus, with a high enough signal-to-noise ratio and a large enough spectral coverage}, the spectral resolution of transmission and emission spectra can translate into vertical resolution of the temperature and composition of the atmosphere. Combining those techniques, EChO will provide a three dimensional vision of numerous close-in planets.

Hundreds of close-in transiting planets with very different gravities and orbital periods are already known and more will be discovered and confirmed before the launch of the mission. Although, for a given star, the irradiation is only function of the distance to the star, the large diversity in exoplanets stellar hosts ensure a good coverage of the rotation period / equilibrium temperature parameter space.  {  As seen in Fig.~\ref{fig::PlotTeq}, the irradiation temperature can vary by a factor $4$ (corresponding to a factor $256$ for the irradiation flux) between planets with similar rotation period but orbiting different stellar types. Planet gravity, for its part, varies by more than two orders of magnitude among known planets, ranging from $\approx 2.5$ to $\approx500\meter\per\second\squared$. The sample of planets EChO will observe thus covers a large area in the irradiation / rotation / planet gravity parameter space, three of the main parameters shaping the atmospheric circulation.}

\begin{figure}[h]
\centering
\includegraphics[width=0.7\linewidth]{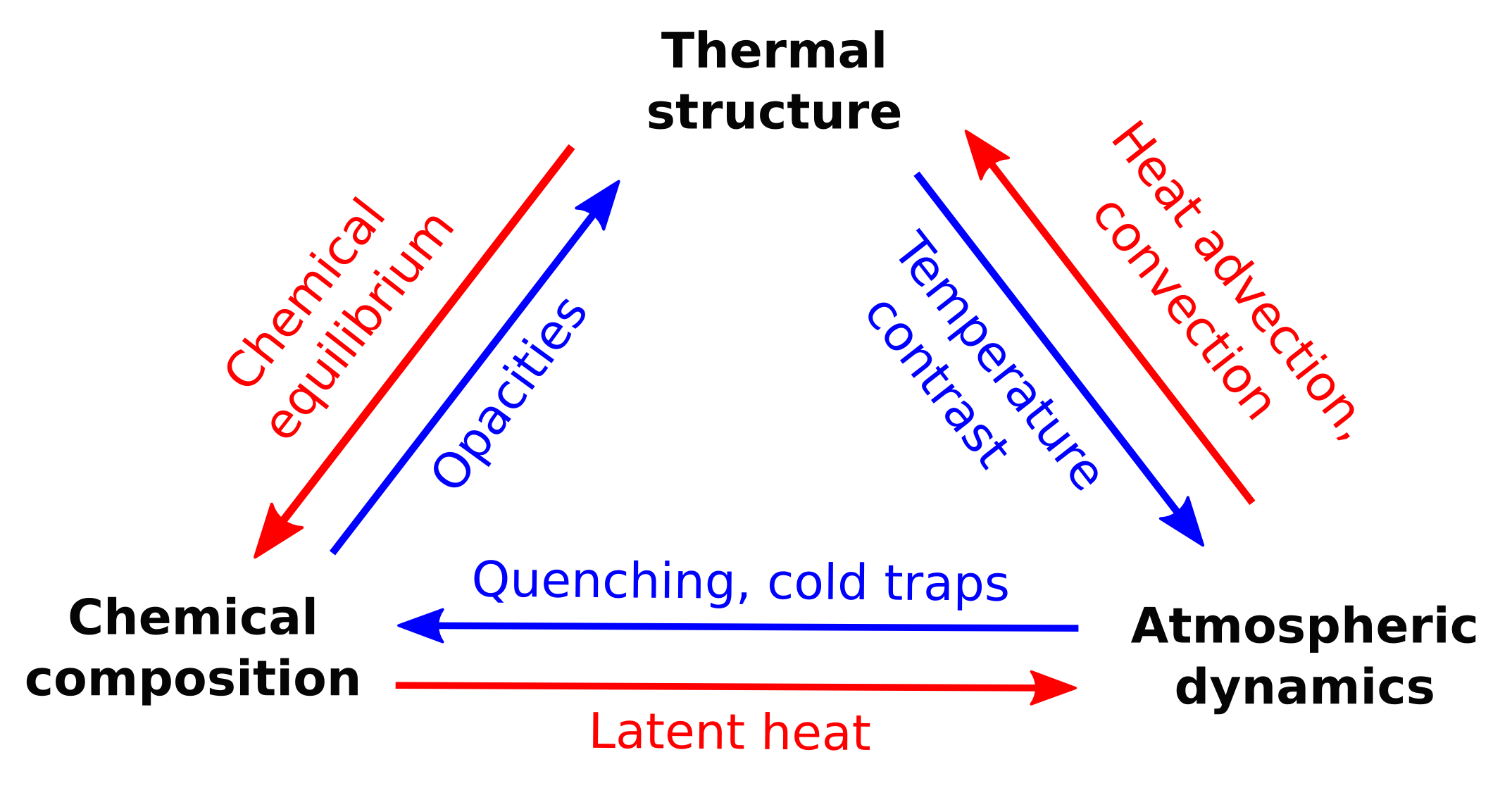}
\captionsetup{justification=centering}
\caption{Schematic view of the main atmospheric characteristics and how they affect each other.}
\label{fig::Triangle}
\end{figure}

Thermal structure, composition and atmospheric circulation are essential characteristics of planetary atmospheres. They affect each other via the different mechanisms described in Fig.~\ref{fig::Triangle}. The thermal structure sets the chemical equilibrium whereas the composition determines the atmospheric opacities, controlling the radiative transfer and thus the temperature. The atmospheric circulation is driven by the temperature contrasts. It transports heat and material, {  which} shapes the temperature and composition both horizontally and vertically. Finally, the presence of ionized material directly affects the circulation via the Lorentz forces. The spatial variation of the temperature and composition, together with their departure from equilibrium are thus signatures of the atmospheric circulation.

EChO can observe hundreds of exoplanets atmospheres with a high spectral resolution and an exquisite photometric precision. It can obtain a full exoplanet spectrum in one observation and will be able to observe periodically a given target. Such a mission is essential to determine the spatio-temporal variability of exoplanets atmospheres and understand their diversity in terms of composition, thermal structure and dynamics. Hereafter we list several key scientific questions concerning {  the thermal structure and atmospheric dynamics of} gas giant atmospheres that EChO's observations will help to solve.  {  Questions related to atmospheric chemical composition are treated in a separate article.}

\section*{Key questions in atmospheric structure and dynamics to be addressed by an EChO-class mission}

\subsection{What is the longitudinal structure of the temperature
in hot Jupiter atmospheres, and how does it depend on depth?} 

\begin{figure}[h!]
\begin{minipage}[c]{0.48\linewidth} 
\includegraphics[height=8.5cm]{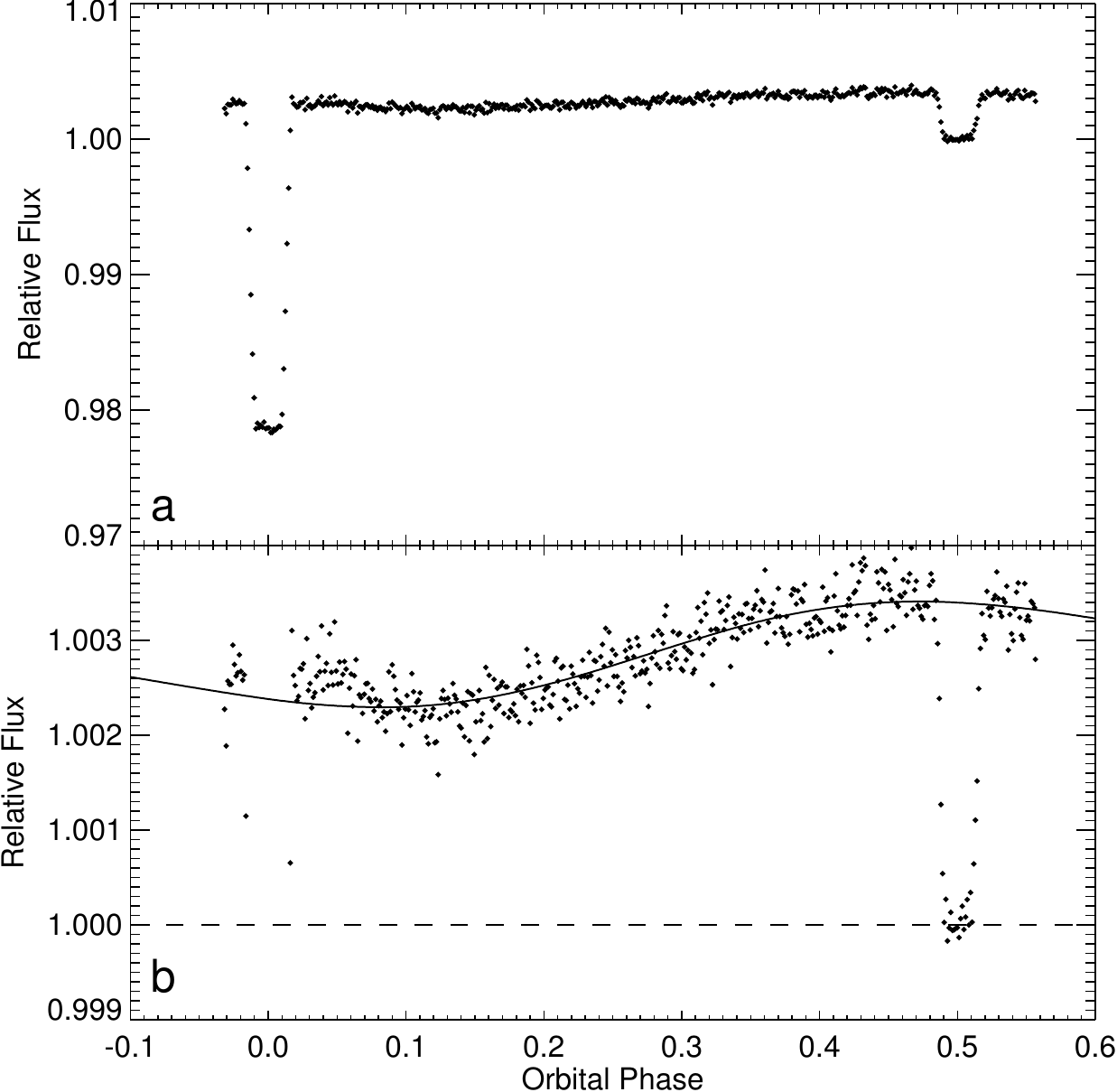}
\centering
  \caption{Thermal phase curve of HD189733 observed with the IRAC instrument on the Spitzer Space Telescope at 8 microns by~\citet{Knutson2007}. In the top panel, the transit (orbital phase 0) and secondary eclipse of the planet orbital phase 0.5) are visible. In the bottom panel, the increase of flux between the transit and the secondary eclipse is due to the planet phase: before and after the transit the planet shows its cold and thus dark nightside whereas before and after the secondary eclipse it shows its warm, and thus luminous, dayside.{\it Reprinted by permission from Macmillan Publishers Ltd. Nature Copyright 2007.} }
\label{fig::Knutson1}
\end{minipage}
\hfill
\begin{minipage}[c]{0.48\linewidth} 
\centering
\includegraphics[height=4.5cm, trim=0 0 40 0]{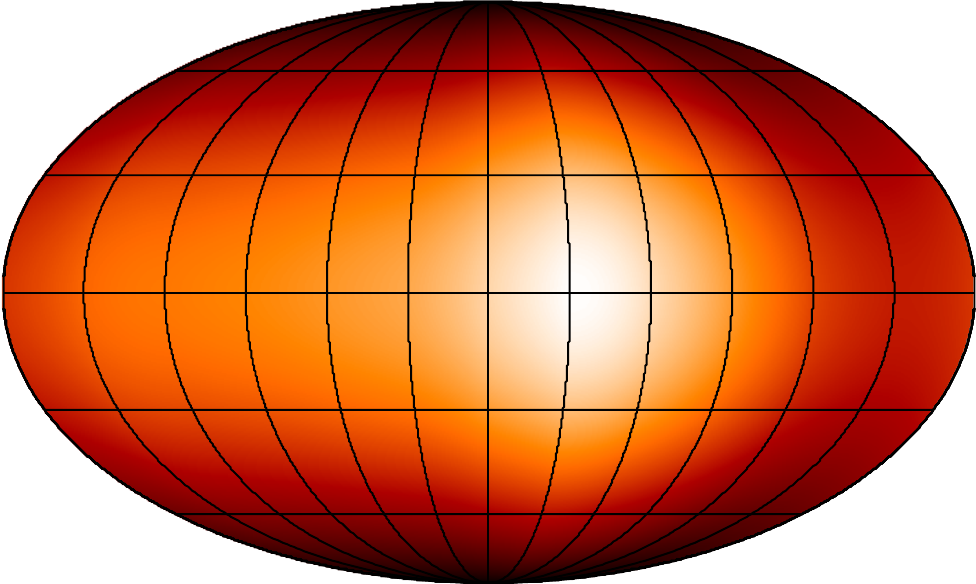}
\includegraphics[height=4cm]{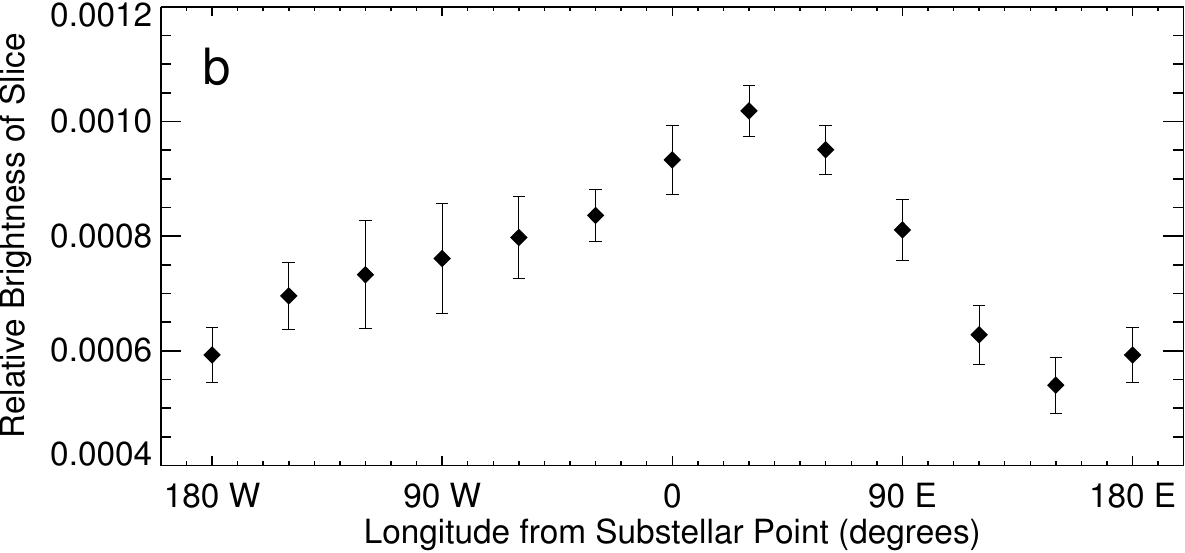}
  \caption{Longitudinal temperature map of the planet HD189733b retrieved from the phase curve observation depicted in the previous figure~\citep[from]{Knutson2007}. The shift of the hottest point of the planet east of the substellar point is attributed to fast eastward equatorial winds~\citep{Showman2009}.{\it Reprinted by permission from Macmillan Publishers Ltd. Nature Copyright 2007.}              
\vspace{1.8cm}}
\label{fig::Knutson2}
\end{minipage}
\end{figure}

High-quality lightcurves---as obtainable from EChO for a wide range of
close-in planets---will allow longitudinal maps of brightness
temperature to be derived.  This will allow the
longitudinal locations of hot and cold spots, among other features, to
be identified; observations at many wavelengths will allow the
depth-dependence to be determined in the range $\sim$0.001--10
bar. Spitzer observations of several hot Jupiters, including HD
189733b~\citep{Knutson2007,Knutson2009,Knutson2012}, Ups And b~\citep{Crossfield2010}, and WASP-43b~\citep{Stevenson2014} indicate that the hottest regions are displaced eastward of
the substellar point by tens of degrees of longitude or more (see Figs.~\ref{fig::Knutson1} and~\ref{fig::Knutson2}).  This
phenomenon was predicted and has now been reproduced in a
wide range of three-dimensional circulation models under conditions
appropriate to benchmark hot Jupiters such as HD 189733b and HD
209458b \citep{Showman2002,Cooper2005,Showman2008,Showman2009,Menou2009,Dobbs-Dixon2008,Dobbs-Dixon2010,Rauscher2010,Rauscher2012b,Heng2011a,Heng2011,Perna2012}. In these models, the eastward
displacement results from advection by an eastward ``superrotating''
jet stream at the equator.  Theory shows that, on tidally locked
planets, such superrotation is the natural result of the day-night
heating pattern, which leads to planetary-scale waves that pump
angular momentum to low latitudes~\citep{Showman2011}. Nevertheless, current predictions---yet to be tested---suggest
that the longitudinal offset of the hotspot should scale inversely
with incident stellar flux~\citep{Showman2011,Perna2012,Showman2013}.  The extent to which such longitudinal
offsets are prevalent on hot Jupiters---and their dependence on
incident stellar flux, planetary rotation rate, atmospheric
composition, and other factors---remains unknown.  {  Recent
magnetohydrodynamic calculations that properly represent the full
coupling of the dynamics to the magnetic field furthermore suggest that,
under particularly hot conditions, a westward equatorial jet can sometimes 
emerge~\citep{Rogers2014,Rogers2014a}, potentially
leading to a {\it westward} hot spot offset in these cases.}
EChO can address
this question with a broad census, {  determining the amplitude and
sign of the offset under a broad range of conditions,} and map the depth dependence of
these features.

\subsection{What sets the day-night temperature contrast?  How
does it vary with depth (wavelength) and among different planets?
What is the mechanism that controls the day-night temperature
contrast on tidally locked planets?}

Current lightcurve observations have allowed the day-night brightness
temperature contrast to be determined for over a dozen hot Jupiters.
These observations suggest a trend wherein cooler planets exhibit
modest fractional day-night temperature contrasts whereas hotter
planets exhibit near-unity fractional day-night temperature variations~\citep{Cowan2011,Perna2012,Perez-Becker2013a}. {  As emphasized by
Perez-Becker \& Showman 2013,} the details of this trend place strong constraints on the
mechanisms that maintain the day-night temperature differences on hot
Jupiters (e.g., on the relative roles of horizontal advection, vertical advection, wave
propagation, and radiative cooling) and on the conditions under which frictional drag and ohmic drag become important~\citep{Li2010,Rauscher2012a,Rauscher2013, Showman2013}. Current observations exist at
only a few broadband wavelengths, and full spectral information as obtainable
from EChO would provide significant information on how the
transition from small to large fractional day-night flux difference
depends on wavelength, and in turn how this transition depends
on depth in the atmosphere.
{  
\subsection{What physical mechanisms determine the vertical temperature profile at the terminator of the planet ?}
The terminator of close-in, tidally locked planets is extremely interesting but very complex. It is located at the middle of the largest temperature gradients and where the fastest winds are present. Hydrodynamics shocks might be present~\citep{Heng2012a}. Scattering should become important due to the grazing path of the stellar rays~\citep{Fortney2005b}. Condensation of numerous species is expected to take place close to the terminator, depositing latent heat and increasing even more the importance of scattering. From the combined effects of the dynamics and the condensation processes, a significant differences in the cloud coverage between the western and the eastern atmospheric limbs is expected~\citep{Iro2005}. Whether the ions produced in the hot dayside recombine before or after crossing the terminator will influence the strength of the magnetic forces acting on the fluid. At low pressures, non local thermodynamic equilibrium (LTE) effects should also play a major role~\citep{Barman2002}.

\begin{figure}[h]
 \includegraphics[width=0.495\linewidth,clip]{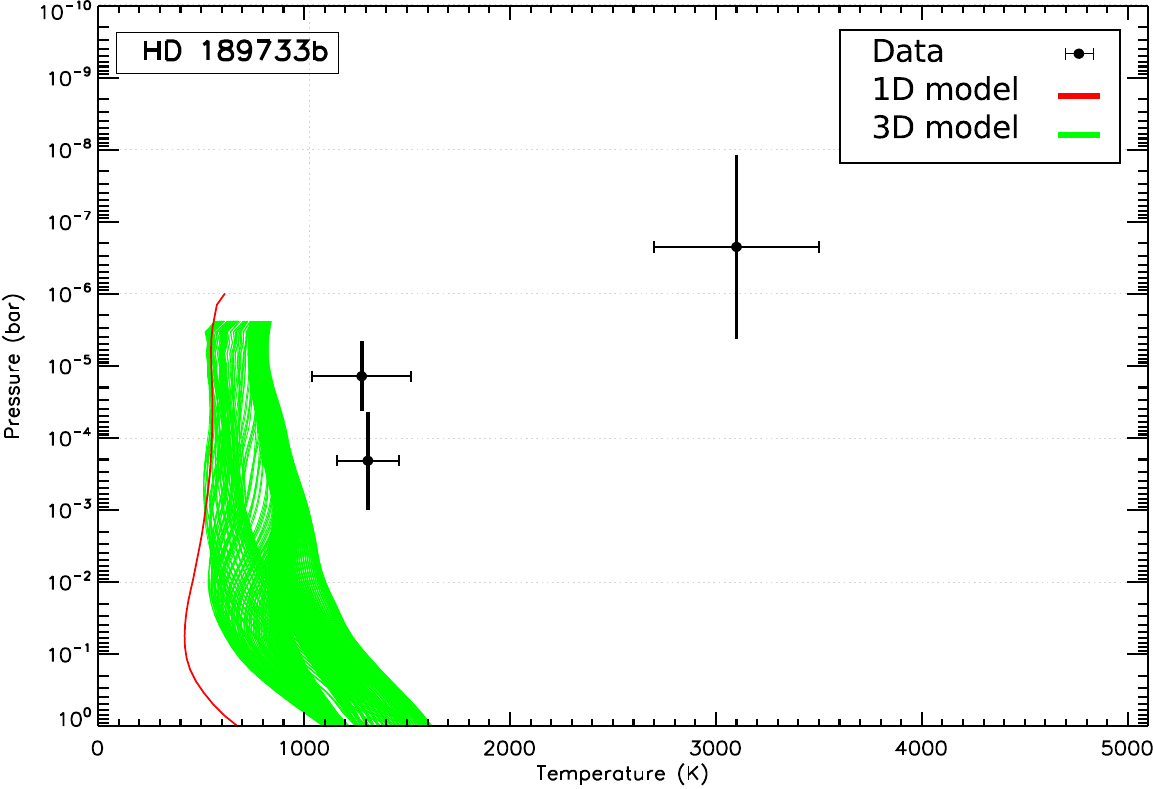}
  \includegraphics[width=0.495\linewidth,clip]{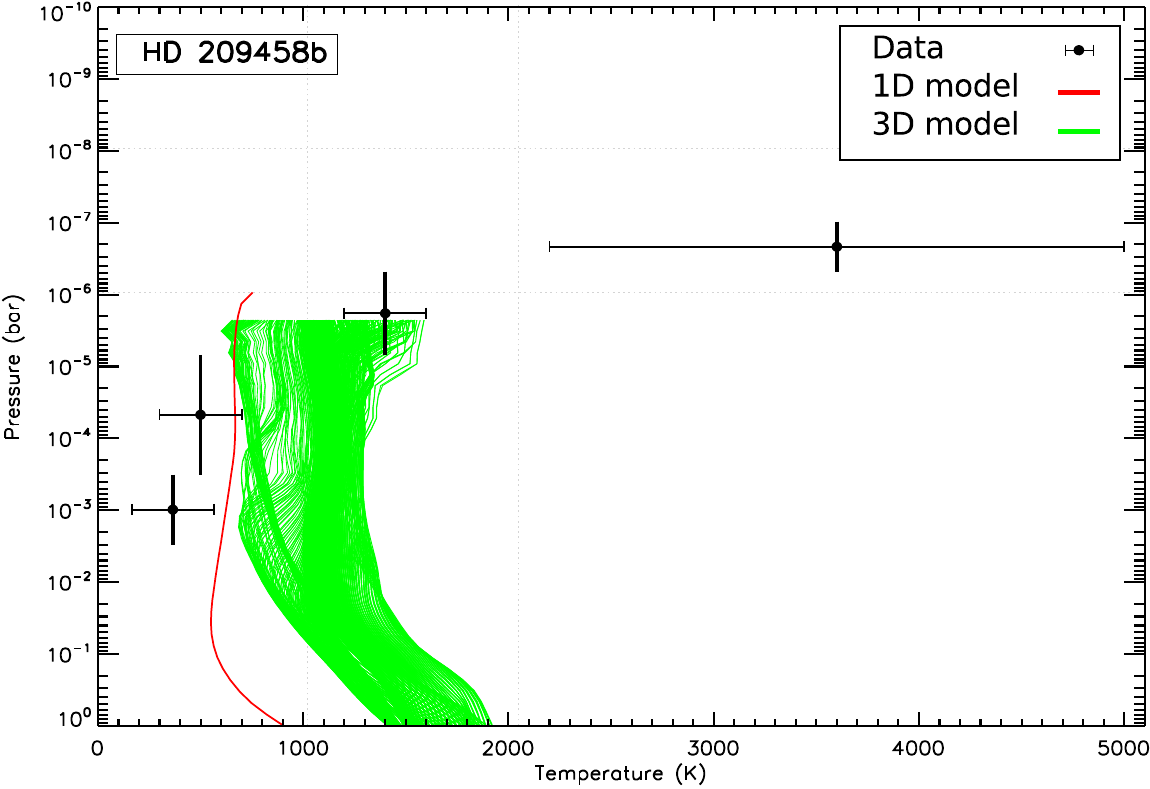}
\caption{Pressure-temperature profile at the terminator of the hot Jupiter HD189733b (left) and HD209458b (right). Data are retrieved from the sodium absorption line of the planet observed during transit by~\citet{Vidal-Madjar2011} and~\citet{Huitson2012}. For HD 209458b, the pressure scale is based on the detection of the Rayleigh scattering by $\rm H_{\rm 2}$. For HD 189733b, the pressure scale is model dependent: it is determined assuming that the top of the cloud deck is at $10^{-4}\,\bbar$. The red line is obtained from the grid of 1D numerical models used in~\citet{Parmentier2014b}. The green lines are all the limb temperature profiles predicted by the 3D model (SPARC/MIT GCM). The difference between the 1D and the 3D temperature profiles is mainly due to the advection of heat by the atmospheric circulation. At pressures lower than $10^{-5}\,\bbar$, non-LTE effects, not taken into account in the models become important~\citet{Barman2002}. Figure adapted from~\citet{Huitson2012}.}
\label{fig::Huitson2012}
\end{figure}

The temperature at the terminator of a planet can be retrieved from the slope of the spectral features apparent in the transit spectrum~\citep{LecavelierDesEtangs2008}. From the absorption feature of the Sodium D line, the temperature profile at the terminator of HD 189733b~\citep{Vidal-Madjar2011} and HD 209458b~\citep{Huitson2012} have been retrieved. As shown in Figure~\ref{fig::Huitson2012}, the retrieved temperatures in the upper atmosphere of HD 189733b and HD 209458b are larger than predicted by current LTE models. They are nonetheless consistent with observations of hot hydrogen in the upper atmosphere of HD 209458b by~\citet{Ballester2007} and necessary to explain the extended atmosphere observed in both planets~\citep{Vidal-Madjar2003,LecavelierDesEtangs2010}. At higher pressures, the temperature of HD 209458b is unexpectedly low and cannot be explained by current 1D and 3D models. Those low temperatures are however consistent with the condensation of sodium at low pressures as shown by~\citet{Sing2008}.

EChO will accurately determine the mean temperature profile at the terminator of a wide range of planets from their transit spectrum. It will disentangle the contributions of the dynamical, chemical and radiative processes shaping the temperature profile at the terminator. For the brightest targets, it will observe the differences between the ingress and the egress of the transit, shedding light on the differences in temperature, chemical composition and cloud coverage between the western and the eastern atmospheric limbs.
}

\subsection{What is the latitudinal structure of the temperature
in hot Jupiter atmospheres?}
The high and low latitudes of a planet differ by the amount of irradiation they receive and by the strength of the Coriolis forces. As a result, in hot Jupiters atmospheric models, the circulation patterns change from a deep super-rotating jet at the equator to a day-to-night circulation at the poles~\citep{Showman2013}. Chemical composition and cloud coverage could follow this trend and be significantly different between the poles and the equator~\citep[see][and Fig.~\ref{fig::Clouds} hereafter]{Parmentier2013}. The secondary eclipse of an exoplanet yields latitudinal information about the temperature structure of its atmosphere. During a secondary eclipse, the planet disappears behind its host star. For non-zero impact parameter, the disappearance and appearance of the planet happen by slices that are tilted with respect to the north/south direction. The ingress and egress of an exoplanet's secondary eclipse can thus allow the construction of full two-dimensional maps of the dayside hemisphere ~\citep{Majeau2012,DeWit2012}, in opposition to phase curves that lead to longitudinal maps only.  Furthermore, as each wavelength probes different optical depth of the dayside atmosphere, multi-wavelength observations, as the ones EChO will provide, can allow tri-dimensional maps of the atmosphere. As an example, the eclipse mapping of HD\, 189733b using Spitzer 8 microns data constrains its hot spot to low latitudes and provides independent confirmation of its eastward shift relative to the substellar point~\citep{Majeau2012,DeWit2012}.

\begin{figure}[h!]
\includegraphics[width=1\linewidth]{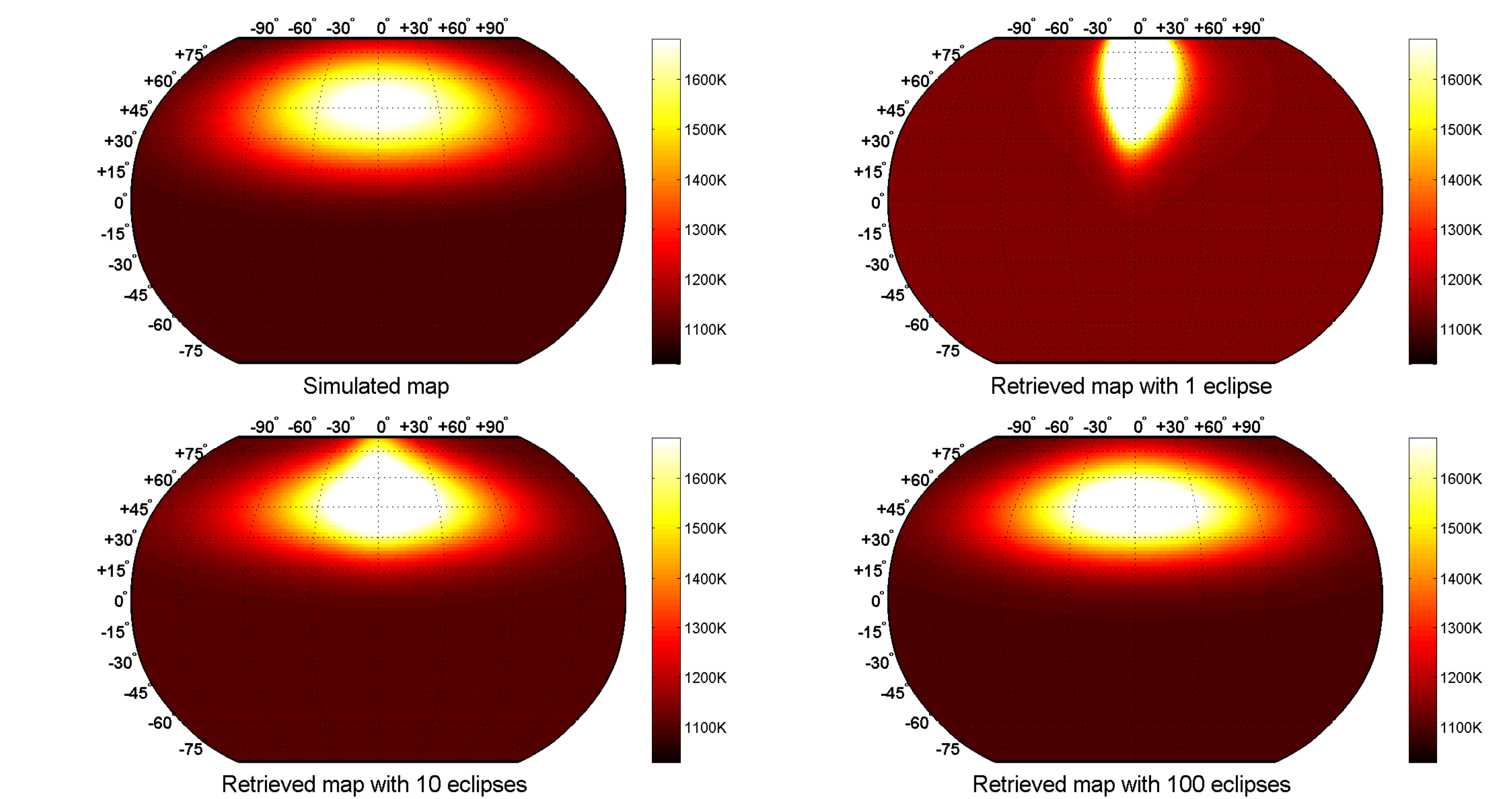}
\centering
  \caption{ {  Simulated retrieval of dayside brightness temperature
      patterns using ingress/egress mapping for a hypothetical case
      where a large thermal hotspot resides in the high northern
      latitudes of the dayside.  Planetary and stellar parameters of
      HD 189733b are adopted.  The top left map depicts the synthetic
      data.  The top right, bottom left, and bottom right shows the
      ability of ingress/egress mapping to recover the temperature
      structure of the synthetic data with 1, 10, and 100 secondary
      eclipses observed by EChO, respectively, in a spectral bin of
      resolution $20$.}}
  \label{fig::maps}
\end{figure}

Based on the technique developed by~\citet{DeWit2012} we present in Fig.~\ref{fig::maps} the map retrieval of a synthetic version of the hot Jupiter HD\,189733b\footnote{ {  We use EChO's noise model introduced in \cite{Barstow2013}. In particular, we use a telescope effective area of $1.13$ square meter, a detector quantum efficiency of $0.7$, a duty-cycle of $0.8$, and an optical throughput of 0.378 from 2.5  to 5 $\mu$m, relevant for this simulation showed in Fig.~\ref{fig::maps}}} with a hypothetical hot spot with a temperature contrast of $\Delta T/T\approx30\%$ located {  in} the northern hemisphere. Such a hot spot in a given spectral bin could be formed by the presence of patchy clouds (see Fig.~\ref{fig::Clouds}) or chemical differences between the poles and the equator. With one secondary eclipse, EChO will detect the presence of latitudinal asymmetry in the planet's brightness distribution.  With $\sim$10 (resp. $\sim$100) secondary eclipses, the temperature contrast will be measured with a precision of $300\,\rm{K}$ (resp. $100\,\rm{K}$) and the latitudinal location of the hot-spot will be known with a precision of $10\degree$ (resp. $3.5\degree$). This observations will be available in different spectral intervals, with a spectral resolution of $\approx20$, for the most favorable targets. 
 
\subsection{How common are clouds, what are they made of, and
what is their spatial distribution?}  
\begin{figure}[h!]
\includegraphics[width=1\linewidth]{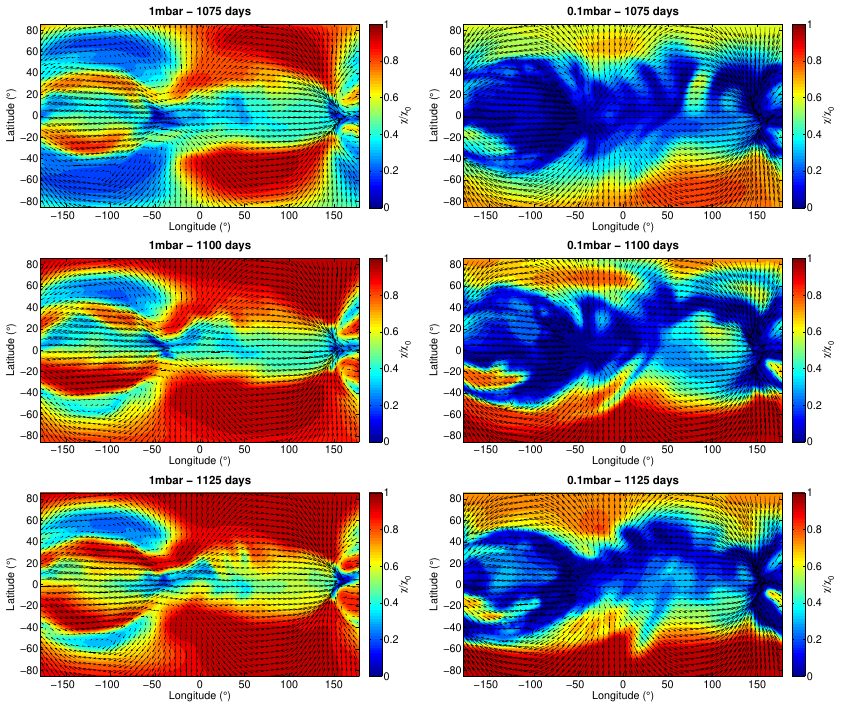}
\centering
  \caption{Spatio-temporal variability of tracer particles (color) and winds (arrows) representing clouds in a hot-Jupiter model of~\citet{Parmentier2013}. The particles efficiently trace the main circulation patterns of the atmosphere.}
  \label{fig::Clouds}
\end{figure}

The atmospheres of many hot and warm
Jupiters have temperatures that cross the condensation curves for
various refractory materials, suggesting that cloud formation may be
an important process on some of those planets. Transmission spectra
indicate that HD 189733b and perhaps HD 209458b exhibit haze-dominated
atmospheres~\citep{Pont2013,Deming2013}.  This may also be
true for the super-Earth GJ 1214b~\citep[e.g.][]{Bean2011,Berta2012,Morley2013} and GJ 3470b~\citep{Crossfield2013,Nascimbeni2013}.  Given the cold conditions on the
nightsides of typical hot Jupiters, many chemical species should
condense on the nightside.  Three-dimensional circulation models including condensable tracers~\citep{Parmentier2013}
 indicate that complex spatial
distributions of clouds---on both the dayside and nightside---can
result from such nightside condensation (see Fig.~\ref{fig::Clouds}).

Multi-wavelength lightcurves obtained by EChO will provide
major constraints not only on the chemical composition and thermal
structure but on the existence and properties of clouds in gas giant's atmospheres. Phase curves in the visible frequency range will provide insight on the longitudinal variation in albedo along the planet, which could be a strong signature of inhomogeneous cloud coverage on the planet atmosphere~\citep{Demory2013,Heng2013}. By monitoring planets with widely different equilibrium temperatures, EChO is expected to characterize the transition from cloudy to cloudless atmospheres and the change in the dominant condensable species with equilibrium temperature, from silicate clouds at high temperatures to water clouds in temperate planets.

{  
\subsection{How common are stratospheres, and what determines their distribution and properties ?}  

Thermal inversions are a natural consequence of visible/UV absorption of the incident star light high in the atmosphere. For an isolated planetary atmosphere in hydrostatic equilibrium and
no local energy sources, the atmospheric temperature decreases
with pressure. In planetary atmospheres irradiated by their
host star, strong optical/UV absorbers in the upper layers can intercept part of the incident star light.
With such a local heating, a zone where the temperature increases with decreasing pressure can form. Most solar system planets have temperature inversions in their atmospheres. In Earth atmosphere it is caused by ozone, which is a strong absorber in the UV~\citep{Chamberlain1987}. In Jupiter, it is mainly caused by the strong absorption in the visible by hazes resulting from methane photochemistery.

The compounds producing thermal inversions in solar system atmospheres do not survive the high temperatures of hottest hot Jupiters. Nevertheless, it has been proposed that thermal inversions in the $\sim1\,\milli\bbar--1\,\bbar$ level could form in the atmosphere of very hot Jupiters due to the strong absorption of the incident stellar radiation in the visible by gaseous titanium oxide~\citep{Hubeny2003}, a compound present in brown dwarfs with similar atmospheric temperatures~\citep{Kirkpatrick2005}. The so-called TiO-hypothesis differentiates between planets hot enough to have gaseous TiO and thus a thermal inversion and planets too cold to have gaseous TiO and thus without thermal inversion~\citep{Fortney2008}. Evidence for the presence of a thermal inversion have been claimed for several planets. Most of these claims were based on the ratio between the $3.6\,\micro\meter$ and the $4.5\,\micro\meter$ thermal fluxes observed with the Spitzer space telescope~\citep{Knutson2008,Burrows2008}. Assuming that water is the main absorber at those wavelengths, a higher flux at $4.5\,\micro\meter$  than at $3.6\,\micro\meter$ can be interpreted as an emission band, created by an inverted temperature profile whereas a smaller flux at $4.5\,\micro\meter$ than at $3.6\,\micro\meter$ can be interpreted as an absorption feature, resulting from a non-inverted temperature profile. Up to now, most of the claims did not survive a more exhaustive analysis that included a large range of possible atmospheric chemical composition and temperature profiles~\citep{Madhusudhan2010}. In current data there is thus no strong evidence for a thermal inversion but it is not ruled out either~\citep{Hansen2014}.

Given the apparent lack of large thermal inversions and strong observational signatures of TiO in the transit spectrum of several planets~\citep[e.g.][]{Desert2008,Huitson2013,Sing2013}, many authors challenged the TiO hypothesis. Condensation in the deep atmosphere~\citet{Showman2009,Spiegel2009} or in the nightside of the planet~\citep{Parmentier2013} could deplete TiO from the dayside atmosphere.~\citet{Knutson2010} noted that TiO could be destroyed by the strong stellar FUV flux, implying that only planets orbiting low activity stars could have an inversion.~\citet{Madhusudhan2011} showed that atmospheres with a carbon to oxygen ratio higher than one should have a reduced TiO abundance, making them unable to maintain a thermal inversion.~\citet{Zahnle2009} and ~\citet{Pont2013} proposed that absorption by hazes instead of TiO could be responsible for the thermal inversion whereas~\citet{Menou2012} showed that ohmic dissipation could also lead to an inverted temperature profile. 

EChO will perform a broad census of which hot Jupiters exhibit
a thermal inversion and which do not, and will determine to which extent the presence of thermal inversions correlates with incident stellar flux,
stellar activity, atmospheric composition, day/night temperature gradients and other parameters.
Because EChO will obtain full IR spectra from which absorption and emission
features can be well identified, the determination of whether
a planet exhibits a stratosphere---and the pressure range of any
stratosphere---will be much more robust than possible with existing
Spitzer and groundbased data. Moreover,
spectral features seen in transit and secondary eclipse will provide
strong constraints on the specific chemical absorber that allows
for the existence of stratospheres.
}

\subsection{What are the main dynamical regimes and what determines
the shift from one to another?}

Hot Neptunes and Jupiters span an enormous range of incident stellar
fluxes, orbital parameters, masses, surface gravities, and rotation
rates, among other parameters.  Not surprisingly, then, theory and
numerical simulations suggest that such planets exhibit several
fundamentally different circulation regimes depending on these
parameters.  Most circulation models to date have emphasized the
benchmark hot Jupiters HD 189733b and HD 209458b~\citep{Showman2002,Cooper2005,Showman2008,Showman2009,Menou2009,Dobbs-Dixon2008,Dobbs-Dixon2010,Thrastarson2010,Thrastarson2011,Rauscher2010,Rauscher2012b,Lewis2010,Heng2011,Heng2011a,Perna2012,Miller-RicciKempton2012,Parmentier2013}.  These models tend to produce several broad zonal
(east-west) jets including a fast superrotating equatorial jet, and
day-night temperature differences of hundreds of Kelvin at photospheric
levels.  Nevertheless, recent theoretical explorations of wider
parameter spaces suggest that at extremely large stellar fluxes, the
fractional day-night temperature differences increases and the
longitudinal offset of hot spots decreases~\citep{Perna2012,Perez-Becker2013a}.  This shift is also accompanied by a
shift from a circulation dominated by zonal (east-west) jets at
moderate stellar flux to a circulation dominated by day-to-night flow
at extreme stellar flux~\citep{Showman2013}.  At orbital separations
beyond those typically identified with hot Jupiters ($>0.1\,$AU),
models suggest that the eastward equatorial jet will give way to a
circulation exhibiting one or more eastward jets in the midlatitudes
of each hemisphere generated by baroclinic instability---a pattern
more reminiscent of Earth or Jupiter~\citep{Showman2012}.  {  The spatial variation of temperature, clouds, and chemical composition can efficiently trace the atmospheric circulation patterns~\citep{Parmentier2013}. By determining those spatial variations for a wide range of planetary conditions, EChO will determine the main circulation regimes of exoplanets atmospheres.} 

\subsection{What is the role of magnetic coupling in the circulation
of hot exoplanets?}

Several authors have suggested that, at the extreme temperatures
achieved on the most highly irradiated hot Jupiters, thermal
ionization may allow a coupling of the atmosphere to the planet's
magnetic field, causing the Lorentz force to become dynamically
important~\citep{Perna2010,Perna2010a,Rauscher2013}. This
could lead to qualitative changes in the day-night temperature
difference and the geometry and speed of the global wind pattern
relative to an otherwise similar planet without such coupling~\citep{Batygin2013}.
Dynamical coupling to the magnetic field could even allow
feedbacks that influence the existence and amplitude of a dayside
stratosphere~\citep{Menou2012}. Moreover, such coupling could lead
to Ohmic dissipation, with possible implications for the planet's
long-term evolution~\citep{Batygin2010,Perna2010a,Huang2012,Wu2013}. The sensitivity of the magnetic effects to the ionisation rate -- given by the composition and the temperature profile -- will allow EChO to identify their role in the hottest planets.

\subsection{Are hot Jupiters temporally variable, and if so,
what is the nature and distribution of the variability?}

Atmospheres of planets in the solar system are turbulent, leading to
temporal fluctuations on a wide range of space and time scales.  This
question is also a crucial one for hot Jupiters, especially because
the temporal behavior of any variability contains telltale clues about
the atmospheric state that would be hard to obtain using other 
techniques.  A variety of searches for variability have taken place
over the years, so far without any firm detections of variability.
Using Spitzer observations of  {  seven secondary-eclipses} of HD 189733b,~\citet{Agol2010} demonstrated an upper limit of 2.7\% of the variability of
the secondary-eclipse depth at 8 $\mu$m.  Most
3D circulation models of typical hot Jupiters exhibit relatively
steady circulation patterns; for example, circulation models coupled
to radiative transfer predict variability in the secondary-eclipse
depth of $\sim$1\% in the Spitzer IRAC bandpasses~\citep{Showman2009}.  Nevertheless, some circulation models predict
high-amplitude variability of up to 10\% or more at global scales~\citep{Cho2003,Cho2008,Rauscher2007}. The amplitude and temporal spectrum of variability have much
to tell about the basic atmospheric structure.  Periods of 
variability are likely to be linked to the periods for dynamical instabilities in the atmosphere.
In turn, these fundamental periods are influenced by the structure of the circulation's basic state including the  stratification (e.g., the Brunt-Vaisala frequency), the vertical shear of the horizontal wind, and other parameters. As a dedicated mission, EChO will be able to observe systematically \emph{all} transits and secondary eclipses of a given planet for a given amount of time and shed light on the different timescale and on the amplitude of the variability of a handfull of hot Jupiters. This way, EChO will allow insights into the dynamics not obtainable in any other way.

\subsection{What are the conditions in the deep, usually unobservable atmosphere ?}

 \begin{SCfigure*}[][h]
\includegraphics[clip=true,width=0.6\linewidth]{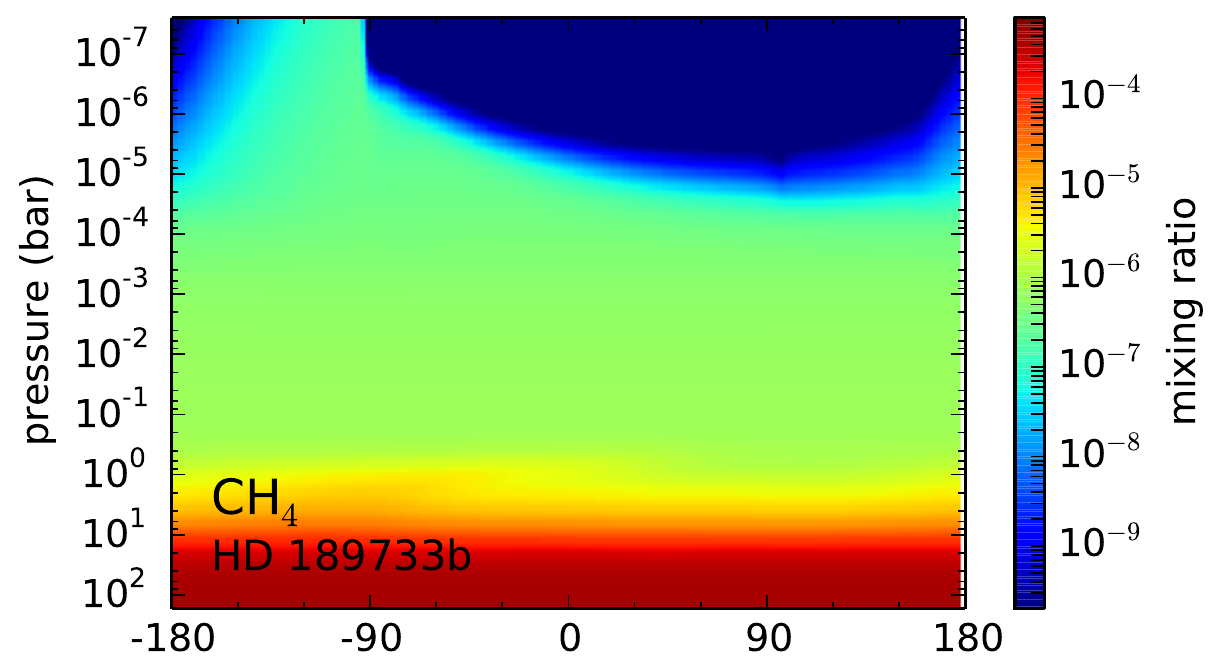}
\centering
  \caption{Abundance of methane in the equatorial plane of HD 189733b predicted by the pseudo-2D chemical model of~\citet{Agundez2014}. The x-axis represents longitude with respect to the substellar point. In the dayside the abundance in the $10^{-5}-0.1\,\bbar$ pressure range is quenched to the abundance at the $0.1\,\bbar$ level. At lower pressures photochemistery becomes important and the abundance drops. the nightside abundance is quenched to the dayside one due to the horizontal advection by an eastward jet.}
  \label{fig::Agundez2014}
\end{SCfigure*}

EChO observations can be used to detect and infer the atmospheric
abundances of major molecules potentially including H$_2$O, CO,
CH$_4$, CO$_2$, and various other trace and/or disequilibrium species~\citep[see][]{Barstow2013}.
To these extent that these species exhibit chemical interactions with
short timescales, they may exhibit spatially variable
three-dimensional distributions (e.g., differing dayside and nightside
abundances).  Any detected spatial variations or homogeneity in such chemical species
across the planet would thus provide important constraints on the
dynamics.

Several species, including CO and CH$_4$, are predicted to have long
interconversion timescales, implying that they will be chemically
``quenched'' in the observable atmosphere at constant abundances that
should vary little from one side of the planet to the other~\citep[see Figure~\ref{fig::Agundez2014} and also][]{Cooper2006,Moses2011,Agundez2014}. The quench level---above which
the abundances are in disequilibrium and below which they are
approximately in equilibrium--is predicted to be at $\sim$0.1--10 bars
pressure on typical hot Jupiters~\citep{Cooper2006,Agundez2014} and
even deeper for cooler planets.
Interestingly, this can be deeper than directly probed by thermal
emission measurements (which sense pressures less than $\sim$10 bar).  
Because the quenched abundances depend
on the atmospheric vertical mixing rate, this implies that precise
measurements of the CO and CH$_4$ abundances will place constraints
on the dynamical mixing rates at pressures deeper than can be
directly sensed.  These insights on the dynamics via chemistry will
thus be highly complementary to insights obtained on the dynamics 
from light curves and ingress/egress mapping. Moreover, they will give constrains on the deep atmosphere, a fundamental zone for understanding the interior and evolution of gas giant planets~\citep{Guillot2002}. On cooler planets,
quenching in the N$_2$/NH$_3$ system can provide analogous insights.

\subsection{Why are some hot Jupiters inflated?}

Transit observations show that many hot Jupiters have radii larger
than can be expected from standard evolution models~\citep[see the review by][]{Guillot2005}. The best way of explaining these
radii is that some hot Jupiters experience an interior heat source
(not accounted for in ``classical'' evolution models) that maintains
a large interior entropy and thereby planetary radius.  Several
explanations have been put forward for this missing energy source,
including tidal dissipation ~\citep[e.g.][]{Bodenheimer2001}, 
mechanical energy transported downward into the interior by the atmosphere~\citep{Guillot2002}, suppression of convective heat loss
in the interior as a result of compositional layering~\citep{Chabrier2007,Leconte2012},
and Ohmic dissipation associated with ionized atmospheric winds
~\citep{Batygin2010,Perna2010,Huang2012,Wu2013}. However, the amount of extra-heating needed to keep hot Jupiters inflated is strongly affected by the ability of the atmosphere to transport the energy from the deep interior to the outer space~\citep{Guillot2011}. The efficiency of this transport is tied to the deep atmospheric temperature and it's spatial variations~\citep{Rauscher2014}. {  The deep temperature is partly determined by the ability of the upper atmosphere to absorb and re-emit the incoming stellar irradiation~\citep{Parmentier2014a,Parmentier2014b}. EChO will determine the chemical composition and the thermal profile of the observable atmosphere. This will restrict the range of possible thermal structures for the deep atmosphere, providing better constraints on the strength of the unknown mechanism inflating hot Jupiters}.

\subsection{How does the circulation respond to seasonal and
extreme forcing?}

Several transiting hot Jupiters, including HD 80606b, HAT-P-2b, and 
HD 17156 have orbital eccentricities exceeding 0.5, which imply
that these planets receive an order of magnitude or more stellar
flux at apoapse than at periapse.  This extreme time-variable heating
may have significant effects on the atmospheric circulation~\citep{Kataria2013}. As the planet goes back and forth between apoapse and periapse, EChO will provide a unique opportunity to see the atmosphere heating up and cooling down at different wavelength, measuring its global thermal inertia~\citep{Lewis2013} and how it varies with depth. Then, those heating and cooling rates can be used to better understand the atmospheric dynamics of planets on a circular orbits, where this measurement is not possible.

\section*{Conclusion}

EChO is a dedicated instrument to observe exoplanets atmospheres proposed to the European Space Agency. {  Although it was not selected in 2014, its exquisite photometric precision and high spectroscopic resolution over a wide spectral range make it the archetype of a future space-mission dedicated to the spectroscopic characterization of exoplanets in tight orbit.} Such a future mission will perform a broad survey of exoplanets atmospheres, exploring a large range of stellar irradiation, rotation period and planetary gravity, three parameters that determine the main dynamical regimes of planetary atmospheres. It will provide a deeper understanding of some benchmark planets, characterizing their three-dimensional thermal, chemical and compositional structure and their variation with time, opening the field of climate study to exoplanets.

\begin{multicols}{2}
\bibliography{./Parmentier2014-ArXiv.bib}

\begin{thebibliography}{100}
\expandafter\ifx\csname natexlab\endcsname\relax\def\natexlab#1{#1}\fi

\bibitem[{{Agol} {et~al.}(2010){Agol}, {Cowan}, {Knutson}, {Deming}, {Steffen},
  {Henry}, \& {Charbonneau}}]{Agol2010}
{Agol}, E., {Cowan}, N.~B., {Knutson}, H.~A., {et~al.} 2010, \apj, 721, 1861

\bibitem[{{Agúndez} {et~al.}(2014){Agúndez}, {Parmentier}, {Venot}, {Hersant},
  \& {Selsis}}]{Agundez2014}
{Agúndez}, M., {Parmentier}, V., {Venot}, O., {Hersant}, F., \& {Selsis}, F.
  2014, A\&A, 564, A73

\bibitem[{{Ballester} {et~al.}(2007){Ballester}, {Sing}, \&
  {Herbert}}]{Ballester2007}
{Ballester}, G.~E., {Sing}, D.~K., \& {Herbert}, F. 2007, \nat, 445, 511

\bibitem[{{Barman} {et~al.}(2002){Barman}, {Hauschildt}, {Schweitzer},
  {Stancil}, {Baron}, \& {Allard}}]{Barman2002}
{Barman}, T.~S., {Hauschildt}, P.~H., {Schweitzer}, A., {et~al.} 2002, \apjl,
  569, L51

\bibitem[{{Barstow} {et~al.}(2013){Barstow}, {Aigrain}, {Irwin}, {Bowles},
  {Fletcher}, \& {Lee}}]{Barstow2013}
{Barstow}, J.~K., {Aigrain}, S., {Irwin}, P.~G.~J., {et~al.} 2013, \mnras, 430,
  1188

\bibitem[{{Batygin} {et~al.}(2013){Batygin}, {Stanley}, \&
  {Stevenson}}]{Batygin2013}
{Batygin}, K., {Stanley}, S., \& {Stevenson}, D.~J. 2013, \apj, 776, 53

\bibitem[{{Batygin} \& {Stevenson}(2010)}]{Batygin2010}
{Batygin}, K. \& {Stevenson}, D.~J. 2010, \apjl, 714, L238

\bibitem[{{Bean} {et~al.}(2011){Bean}, {D{\'e}sert}, {Kabath}, {Stalder},
  {Seager}, {Miller-Ricci Kempton}, {Berta}, {Homeier}, {Walsh}, \&
  {Seifahrt}}]{Bean2011}
{Bean}, J.~L., {D{\'e}sert}, J.-M., {Kabath}, P., {et~al.} 2011, \apj, 743, 92

\bibitem[{{Berta} {et~al.}(2012){Berta}, {Charbonneau}, {D{\'e}sert},
  {Miller-Ricci Kempton}, {McCullough}, {Burke}, {Fortney}, {Irwin}, {Nutzman},
  \& {Homeier}}]{Berta2012}
{Berta}, Z.~K., {Charbonneau}, D., {D{\'e}sert}, J.-M., {et~al.} 2012, \apj,
  747, 35

\bibitem[{{Bodenheimer} {et~al.}(2001){Bodenheimer}, {Lin}, \&
  {Mardling}}]{Bodenheimer2001}
{Bodenheimer}, P., {Lin}, D.~N.~C., \& {Mardling}, R.~A. 2001, \apj, 548, 466

\bibitem[{{Burrows} {et~al.}(2008){Burrows}, {Budaj}, \&
  {Hubeny}}]{Burrows2008}
{Burrows}, A., {Budaj}, J., \& {Hubeny}, I. 2008, \apj, 678, 1436

\bibitem[{{Chabrier} \& {Baraffe}(2007)}]{Chabrier2007}
{Chabrier}, G. \& {Baraffe}, I. 2007, \apjl, 661, L81

\bibitem[{{Chamberlain} \& {Hunten}(1987)}]{Chamberlain1987}
{Chamberlain}, J.~W. \& {Hunten}, D.~M. 1987, {Theory of planetary atmospheres.
  An introduction to their physics andchemistry.}

\bibitem[{{Cho} {et~al.}(2003){Cho}, {Menou}, {Hansen}, \& {Seager}}]{Cho2003}
{Cho}, J.~Y.-K., {Menou}, K., {Hansen}, B.~M.~S., \& {Seager}, S. 2003, \apjl,
  587, L117

\bibitem[{{Cho} {et~al.}(2008){Cho}, {Menou}, {Hansen}, \& {Seager}}]{Cho2008}
{Cho}, J.~Y.-K., {Menou}, K., {Hansen}, B.~M.~S., \& {Seager}, S. 2008, \apj,
  675, 817

\bibitem[{{Cooper} \& {Showman}(2005)}]{Cooper2005}
{Cooper}, C.~S. \& {Showman}, A.~P. 2005, \apjl, 629, L45

\bibitem[{{Cooper} \& {Showman}(2006)}]{Cooper2006}
{Cooper}, C.~S. \& {Showman}, A.~P. 2006, \apj, 649, 1048

\bibitem[{{Cowan} \& {Agol}(2011)}]{Cowan2011}
{Cowan}, N.~B. \& {Agol}. 2011, \apj, 729, 54

\bibitem[{{Crossfield} {et~al.}(2013){Crossfield}, {Barman}, {Hansen}, \&
  {Howard}}]{Crossfield2013}
{Crossfield}, I.~J.~M., {Barman}, T., {Hansen}, B.~M.~S., \& {Howard}, A.~W.
  2013, \aap, 559, A33

\bibitem[{{Crossfield} {et~al.}(2010){Crossfield}, {Hansen}, {Harrington},
  {Cho}, {Deming}, {Menou}, \& {Seager}}]{Crossfield2010}
{Crossfield}, I.~J.~M., {Hansen}, B.~M.~S., {Harrington}, J., {et~al.} 2010,
  \apj, 723, 1436

\bibitem[{{de Wit} {et~al.}(2012){de Wit}, {Gillon}, {Demory}, \&
  {Seager}}]{DeWit2012}
{de Wit}, J., {Gillon}, M., {Demory}, B.-O., \& {Seager}, S. 2012, \aap, 548,
  A128

\bibitem[{{de Wit} \& {Seager}(2013)}]{deWit2013}
{de Wit}, J. \& {Seager}, S. 2013, Science, 342, 1473

\bibitem[{{Deming} {et~al.}(2013){Deming}, {Wilkins}, {McCullough}, {Burrows},
  {Fortney}, {Agol}, {Dobbs-Dixon}, {Madhusudhan}, {Crouzet}, {Desert},
  {Gilliland}, {Haynes}, {Knutson}, {Line}, {Magic}, {Mandell}, {Ranjan},
  {Charbonneau}, {Clampin}, {Seager}, \& {Showman}}]{Deming2013}
{Deming}, D., {Wilkins}, A., {McCullough}, P., {et~al.} 2013, \apj, 774, 95

\bibitem[{{Demory} {et~al.}(2013){Demory}, {de Wit}, {Lewis}, {Fortney},
  {Zsom}, {Seager}, {Knutson}, {Heng}, {Madhusudhan}, {Gillon}, {Barclay},
  {Desert}, {Parmentier}, \& {Cowan}}]{Demory2013}
{Demory}, B.-O., {de Wit}, J., {Lewis}, N., {et~al.} 2013, \apjl, 776, L25

\bibitem[{{D{\'e}sert} {et~al.}(2008){D{\'e}sert}, {Vidal-Madjar}, {Lecavelier
  Des Etangs}, {Sing}, {Ehrenreich}, {H{\'e}brard}, \& {Ferlet}}]{Desert2008}
{D{\'e}sert}, J.-M., {Vidal-Madjar}, A., {Lecavelier Des Etangs}, A., {et~al.}
  2008, \aap, 492, 585

\bibitem[{{Dobbs-Dixon} {et~al.}(2010){Dobbs-Dixon}, {Cumming}, \&
  {Lin}}]{Dobbs-Dixon2010}
{Dobbs-Dixon}, I., {Cumming}, A., \& {Lin}, D.~N.~C. 2010, \apj, 710, 1395

\bibitem[{{Dobbs-Dixon} \& {Lin}(2008)}]{Dobbs-Dixon2008}
{Dobbs-Dixon}, I. \& {Lin}, D.~N.~C. 2008, \apj, 673, 513

\bibitem[{{Ferraz-Mello}(2013)}]{Ferraz-Mello2013}
{Ferraz-Mello}, S. 2013, Celestial Mechanics and Dynamical Astronomy, 116, 109

\bibitem[{{Fortney}(2005)}]{Fortney2005b}
{Fortney}, J.~J. 2005, \mnras, 364, 649

\bibitem[{{Fortney} {et~al.}(2008){Fortney}, {Lodders}, {Marley}, \&
  {Freedman}}]{Fortney2008}
{Fortney}, J.~J., {Lodders}, K., {Marley}, M.~S., \& {Freedman}, R.~S. 2008,
  \apj, 678, 1419

\bibitem[{{Guillot}(2005)}]{Guillot2005}
{Guillot}, T. 2005, Annual Review of Earth and Planetary Sciences, 33, 493

\bibitem[{{Guillot} {et~al.}(1996){Guillot}, {Burrows}, {Hubbard}, {Lunine}, \&
  {Saumon}}]{Guillot1996}
{Guillot}, T., {Burrows}, A., {Hubbard}, W.~B., {Lunine}, J.~I., \& {Saumon},
  D. 1996, \apjl, 459, L35

\bibitem[{{Guillot} \& {Havel}(2011)}]{Guillot2011}
{Guillot}, T. \& {Havel}, M. 2011, \aap, 527, A20

\bibitem[{{Guillot} \& {Showman}(2002)}]{Guillot2002}
{Guillot}, T. \& {Showman}, A.~P. 2002, \aap, 385, 156

\bibitem[{{Hansen} {et~al.}(2014){Hansen}, {Schwartz}, \& {Cowan}}]{Hansen2014}
{Hansen}, C.~J., {Schwartz}, J.~C., \& {Cowan}, N.~B. 2014, ArXiv e-prints

\bibitem[{{Heng}(2012)}]{Heng2012a}
{Heng}, K. 2012, \apjl, 761, L1

\bibitem[{{Heng} \& {Demory}(2013)}]{Heng2013}
{Heng}, K. \& {Demory}, B.-O. 2013, \apj, 777, 100

\bibitem[{{Heng} {et~al.}(2011{\natexlab{a}}){Heng}, {Frierson}, \&
  {Phillipps}}]{Heng2011a}
{Heng}, K., {Frierson}, D.~M.~W., \& {Phillipps}, P.~J. 2011{\natexlab{a}},
  \mnras, 418, 2669

\bibitem[{{Heng} {et~al.}(2011{\natexlab{b}}){Heng}, {Menou}, \&
  {Phillipps}}]{Heng2011}
{Heng}, K., {Menou}, K., \& {Phillipps}, P.~J. 2011{\natexlab{b}}, \mnras, 413,
  2380

\bibitem[{{Huang} \& {Cumming}(2012)}]{Huang2012}
{Huang}, X. \& {Cumming}, A. 2012, \apj, 757, 47

\bibitem[{{Hubeny} {et~al.}(2003){Hubeny}, {Burrows}, \&
  {Sudarsky}}]{Hubeny2003}
{Hubeny}, I., {Burrows}, A., \& {Sudarsky}, D. 2003, \apj, 594, 1011

\bibitem[{{Huitson} {et~al.}(2013){Huitson}, {Sing}, {Pont}, {Fortney},
  {Burrows}, {Wilson}, {Ballester}, {Nikolov}, {Gibson}, {Deming}, {Aigrain},
  {Evans}, {Henry}, {Lecavelier des Etangs}, {Showman}, {Vidal-Madjar}, \&
  {Zahnle}}]{Huitson2013}
{Huitson}, C.~M., {Sing}, D.~K., {Pont}, F., {et~al.} 2013, \mnras, 434, 3252

\bibitem[{{Huitson} {et~al.}(2012){Huitson}, {Sing}, {Vidal-Madjar},
  {Ballester}, {Lecavelier des Etangs}, {D{\'e}sert}, \& {Pont}}]{Huitson2012}
{Huitson}, C.~M., {Sing}, D.~K., {Vidal-Madjar}, A., {et~al.} 2012, \mnras,
  422, 2477

\bibitem[{{Iro} {et~al.}(2005){Iro}, {B{\'e}zard}, \& {Guillot}}]{Iro2005}
{Iro}, N., {B{\'e}zard}, B., \& {Guillot}, T. 2005, \aap, 436, 719

\bibitem[{{Kataria} {et~al.}(2013){Kataria}, {Showman}, {Lewis}, {Fortney},
  {Marley}, \& {Freedman}}]{Kataria2013}
{Kataria}, T., {Showman}, A.~P., {Lewis}, N.~K., {et~al.} 2013, \apj, 767, 76

\bibitem[{{Kirkpatrick}(2005)}]{Kirkpatrick2005}
{Kirkpatrick}, J.~D. 2005, \araa, 43, 195

\bibitem[{{Knutson} {et~al.}(2008){Knutson}, {Charbonneau}, {Allen}, {Burrows},
  \& {Megeath}}]{Knutson2008}
{Knutson}, H.~A., {Charbonneau}, D., {Allen}, L.~E., {Burrows}, A., \&
  {Megeath}, S.~T. 2008, \apj, 673, 526

\bibitem[{{Knutson} {et~al.}(2007){Knutson}, {Charbonneau}, {Allen}, {Fortney},
  {Agol}, {Cowan}, {Showman}, {Cooper}, \& {Megeath}}]{Knutson2007}
{Knutson}, H.~A., {Charbonneau}, D., {Allen}, L.~E., {et~al.} 2007, \nat, 447,
  183

\bibitem[{{Knutson} {et~al.}(2009){Knutson}, {Charbonneau}, {Cowan}, {Fortney},
  {Showman}, {Agol}, {Henry}, {Everett}, \& {Allen}}]{Knutson2009}
{Knutson}, H.~A., {Charbonneau}, D., {Cowan}, N.~B., {et~al.} 2009, \apj, 690,
  822

\bibitem[{{Knutson} {et~al.}(2010){Knutson}, {Howard}, \&
  {Isaacson}}]{Knutson2010}
{Knutson}, H.~A., {Howard}, A.~W., \& {Isaacson}, H. 2010, \apj, 720, 1569

\bibitem[{{Knutson} {et~al.}(2012){Knutson}, {Lewis}, {Fortney}, {Burrows},
  {Showman}, {Cowan}, {Agol}, {Aigrain}, {Charbonneau}, {Deming}, {D{\'e}sert},
  {Henry}, {Langton}, \& {Laughlin}}]{Knutson2012}
{Knutson}, H.~A., {Lewis}, N., {Fortney}, J.~J., {et~al.} 2012, \apj, 754, 22

\bibitem[{{Lecavelier Des Etangs} {et~al.}(2010){Lecavelier Des Etangs},
  {Ehrenreich}, {Vidal-Madjar}, {Ballester}, {D{\'e}sert}, {Ferlet},
  {H{\'e}brard}, {Sing}, {Tchakoumegni}, \& {Udry}}]{LecavelierDesEtangs2010}
{Lecavelier Des Etangs}, A., {Ehrenreich}, D., {Vidal-Madjar}, A., {et~al.}
  2010, \aap, 514, A72

\bibitem[{{Lecavelier Des Etangs} {et~al.}(2008){Lecavelier Des Etangs},
  {Pont}, {Vidal-Madjar}, \& {Sing}}]{LecavelierDesEtangs2008}
{Lecavelier Des Etangs}, A., {Pont}, F., {Vidal-Madjar}, A., \& {Sing}, D.
  2008, \aap, 481, L83

\bibitem[{{Leconte} \& {Chabrier}(2012)}]{Leconte2012}
{Leconte}, J. \& {Chabrier}, G. 2012, \aap, 540, A20

\bibitem[{{Lewis} {et~al.}(2013){Lewis}, {Knutson}, {Showman}, {Cowan},
  {Laughlin}, {Burrows}, {Deming}, {Crepp}, {Mighell}, {Agol}, {Bakos},
  {Charbonneau}, {D{\'e}sert}, {Fischer}, {Fortney}, {Hartman}, {Hinkley},
  {Howard}, {Johnson}, {Kao}, {Langton}, \& {Marcy}}]{Lewis2013}
{Lewis}, N.~K., {Knutson}, H.~A., {Showman}, A.~P., {et~al.} 2013, \apj, 766,
  95

\bibitem[{{Lewis} {et~al.}(2010){Lewis}, {Showman}, {Fortney}, {Marley},
  {Freedman}, \& {Lodders}}]{Lewis2010}
{Lewis}, N.~K., {Showman}, A.~P., {Fortney}, J.~J., {et~al.} 2010, \apj, 720,
  344

\bibitem[{{Li} \& {Goodman}(2010)}]{Li2010}
{Li}, J. \& {Goodman}, J. 2010, \apj, 725, 1146

\bibitem[{{Lubow} {et~al.}(1997){Lubow}, {Tout}, \& {Livio}}]{Lubow1997}
{Lubow}, S.~H., {Tout}, C.~A., \& {Livio}, M. 1997, \apj, 484, 866

\bibitem[{{Madhusudhan} {et~al.}(2011){Madhusudhan}, {Mousis}, {Johnson}, \&
  {Lunine}}]{Madhusudhan2011}
{Madhusudhan}, N., {Mousis}, O., {Johnson}, T.~V., \& {Lunine}, J.~I. 2011,
  \apj, 743, 191

\bibitem[{{Madhusudhan} \& {Seager}(2010)}]{Madhusudhan2010}
{Madhusudhan}, N. \& {Seager}, S. 2010, \apj, 725, 261

\bibitem[{{Majeau} {et~al.}(2012){Majeau}, {Agol}, \& {Cowan}}]{Majeau2012}
{Majeau}, C., {Agol}, E., \& {Cowan}, N.~B. 2012, \apjl, 747, L20

\bibitem[{{Menou}(2012)}]{Menou2012}
{Menou}, K. 2012, \apjl, 754, L9

\bibitem[{{Menou} \& {Rauscher}(2009)}]{Menou2009}
{Menou}, K. \& {Rauscher}, E. 2009, \apj, 700, 887

\bibitem[{{Miller-Ricci Kempton} \& {Rauscher}(2012)}]{Miller-RicciKempton2012}
{Miller-Ricci Kempton}, E. \& {Rauscher}, E. 2012, \apj, 751, 117

\bibitem[{{Morley} {et~al.}(2013){Morley}, {Fortney}, {Kempton}, {Marley},
  {Vissher}, \& {Zahnle}}]{Morley2013}
{Morley}, C.~V., {Fortney}, J.~J., {Kempton}, E.~M.-R., {et~al.} 2013, \apj,
  775, 33

\bibitem[{{Moses} {et~al.}(2011){Moses}, {Visscher}, {Fortney}, {Showman},
  {Lewis}, {Griffith}, {Klippenstein}, {Shabram}, {Friedson}, {Marley}, \&
  {Freedman}}]{Moses2011}
{Moses}, J.~I., {Visscher}, C., {Fortney}, J.~J., {et~al.} 2011, \apj, 737, 15

\bibitem[{{Nascimbeni} {et~al.}(2013){Nascimbeni}, {Piotto}, {Pagano},
  {Scandariato}, {Sani}, \& {Fumana}}]{Nascimbeni2013}
{Nascimbeni}, V., {Piotto}, G., {Pagano}, I., {et~al.} 2013, \aap, 559, A32

\bibitem[{{Parmentier} \& {Guillot}(2014)}]{Parmentier2014a}
{Parmentier}, V. \& {Guillot}, T. 2014, \aap, 562, A133

\bibitem[{{Parmentier} {et~al.}(2014){Parmentier}, {Guillot}, {Fortney}, \&
  {Marley}}]{Parmentier2014b}
{Parmentier}, V., {Guillot}, T., {Fortney}, J.~J., \& {Marley}, M.~S. 2014,
  ArXiv:1311.6322

\bibitem[{{Parmentier} {et~al.}(2013){Parmentier}, {Showman}, \&
  {Lian}}]{Parmentier2013}
{Parmentier}, V., {Showman}, A.~P., \& {Lian}, Y. 2013, \aap, 558, A91

\bibitem[{{Perez-Becker} \& {Showman}(2013)}]{Perez-Becker2013a}
{Perez-Becker}, D. \& {Showman}, A.~P. 2013, \apj, 776, 134

\bibitem[{{Perna} {et~al.}(2012){Perna}, {Heng}, \& {Pont}}]{Perna2012}
{Perna}, R., {Heng}, K., \& {Pont}, F. 2012, \apj, 751, 59

\bibitem[{{Perna} {et~al.}(2010{\natexlab{a}}){Perna}, {Menou}, \&
  {Rauscher}}]{Perna2010}
{Perna}, R., {Menou}, K., \& {Rauscher}, E. 2010{\natexlab{a}}, \apj, 719, 1421

\bibitem[{{Perna} {et~al.}(2010{\natexlab{b}}){Perna}, {Menou}, \&
  {Rauscher}}]{Perna2010a}
{Perna}, R., {Menou}, K., \& {Rauscher}, E. 2010{\natexlab{b}}, \apj, 724, 313

\bibitem[{{Pont} {et~al.}(2013){Pont}, {Sing}, {Gibson}, {Aigrain}, {Henry}, \&
  {Husnoo}}]{Pont2013}
{Pont}, F., {Sing}, D.~K., {Gibson}, N.~P., {et~al.} 2013, \mnras

\bibitem[{{Rauscher} \& {Menou}(2010)}]{Rauscher2010}
{Rauscher}, E. \& {Menou}, K. 2010, \apj, 714, 1334

\bibitem[{{Rauscher} \& {Menou}(2012{\natexlab{a}})}]{Rauscher2012b}
{Rauscher}, E. \& {Menou}, K. 2012{\natexlab{a}}, \apj, 750, 96

\bibitem[{{Rauscher} \& {Menou}(2012{\natexlab{b}})}]{Rauscher2012a}
{Rauscher}, E. \& {Menou}, K. 2012{\natexlab{b}}, \apj, 745, 78

\bibitem[{{Rauscher} \& {Menou}(2013)}]{Rauscher2013}
{Rauscher}, E. \& {Menou}, K. 2013, \apj, 764, 103

\bibitem[{{Rauscher} {et~al.}(2007){Rauscher}, {Menou}, {Seager}, {Deming},
  {Cho}, \& {Hansen}}]{Rauscher2007}
{Rauscher}, E., {Menou}, K., {Seager}, S., {et~al.} 2007, \apj, 664, 1199

\bibitem[{{Rauscher} \& {Showman}(2014)}]{Rauscher2014}
{Rauscher}, E. \& {Showman}, A.~P. 2014, \apj, 784, 160

\bibitem[{{Rogers} \& {Komacek}(2014)}]{Rogers2014a}
{Rogers}, T.~M. \& {Komacek}, T.~D. 2014, Submitted to \apjl

\bibitem[{{Rogers} \& {Showman}(2014)}]{Rogers2014}
{Rogers}, T.~M. \& {Showman}, A.~P. 2014, \apjl, 782, L4

\bibitem[{{Showman} {et~al.}(2008){Showman}, {Cooper}, {Fortney}, \&
  {Marley}}]{Showman2008}
{Showman}, A.~P., {Cooper}, C.~S., {Fortney}, J.~J., \& {Marley}, M.~S. 2008,
  \apj, 682, 559

\bibitem[{{Showman} {et~al.}(2013){Showman}, {Fortney}, {Lewis}, \&
  {Shabram}}]{Showman2013}
{Showman}, A.~P., {Fortney}, J.~J., {Lewis}, N.~K., \& {Shabram}, M. 2013,
  \apj, 762, 24

\bibitem[{{Showman} {et~al.}(2009){Showman}, {Fortney}, {Lian}, {Marley},
  {Freedman}, {Knutson}, \& {Charbonneau}}]{Showman2009}
{Showman}, A.~P., {Fortney}, J.~J., {Lian}, Y., {et~al.} 2009, \apj, 699, 564

\bibitem[{{Showman} \& {Guillot}(2002)}]{Showman2002}
{Showman}, A.~P. \& {Guillot}, T. 2002, \aap, 385, 166

\bibitem[{{Showman} \& {Polvani}(2011)}]{Showman2011}
{Showman}, A.~P. \& {Polvani}, L.~M. 2011, \apj, 738, 71

\bibitem[{{Showman} {et~al.}(2012){Showman}, {Wordsworth}, \&
  {Merlis}}]{Showman2012}
{Showman}, A.~P., {Wordsworth}, R.~D., \& {Merlis}, T.~M. 2012, LPI
  Contributions, 1675, 8090

\bibitem[{{Sing} {et~al.}(2013){Sing}, {Lecavelier des Etangs}, {Fortney},
  {Burrows}, {Pont}, {Wakeford}, {Ballester}, {Nikolov}, {Henry}, {Aigrain},
  {Deming}, {Evans}, {Gibson}, {Huitson}, {Knutson}, {Showman}, {Vidal-Madjar},
  {Wilson}, {Williamson}, \& {Zahnle}}]{Sing2013}
{Sing}, D.~K., {Lecavelier des Etangs}, A., {Fortney}, J.~J., {et~al.} 2013,
  \mnras

\bibitem[{{Sing} {et~al.}(2008){Sing}, {Vidal-Madjar}, {Lecavelier des Etangs},
  {D{\'e}sert}, {Ballester}, \& {Ehrenreich}}]{Sing2008}
{Sing}, D.~K., {Vidal-Madjar}, A., {Lecavelier des Etangs}, A., {et~al.} 2008,
  \apj, 686, 667

\bibitem[{{Spiegel} {et~al.}(2009){Spiegel}, {Silverio}, \&
  {Burrows}}]{Spiegel2009}
{Spiegel}, D.~S., {Silverio}, K., \& {Burrows}, A. 2009, \apj, 699, 1487

\bibitem[{{Stevenson}(2014)}]{Stevenson2014}
{Stevenson}, K. 2014, in Talk at the Exoclimes III conference

\bibitem[{{Thrastarson} \& {Cho}(2010)}]{Thrastarson2010}
{Thrastarson}, H.~T. \& {Cho}, J.~Y. 2010, \apj, 716, 144

\bibitem[{{Thrastarson} \& {Cho}(2011)}]{Thrastarson2011}
{Thrastarson}, H.~T. \& {Cho}, J.~Y. 2011, \apj, 729, 117

\bibitem[{{Tinetti} {et~al.}(2012){Tinetti}, {Beaulieu}, {Henning}, {Meyer},
  {Micela}, {Ribas}, {Stam}, {Swain}, {Krause}, {Ollivier}, {Pace}, {Swinyard},
  {Aylward}, {van Boekel}, {Coradini}, {Encrenaz}, {Snellen},
  {Zapatero-Osorio}, {Bouwman}, {Cho}, {Coud{\'e} de Foresto}, {Guillot},
  {Lopez-Morales}, {Mueller-Wodarg}, {Palle}, {Selsis}, {Sozzetti}, {Ade},
  {Achilleos}, {Adriani}, {Agnor}, {Afonso}, {Allende Prieto}, {Bakos},
  {Barber}, {Barlow}, {Batista}, {Bernath}, {B{\'e}zard}, {Bord{\'e}}, {Brown},
  {Cassan}, {Cavarroc}, {Ciaravella}, {Cockell}, {Coustenis}, {Danielski},
  {Decin}, {De Kok}, {Demangeon}, {Deroo}, {Doel}, {Drossart}, {Fletcher},
  {Focardi}, {Forget}, {Fossey}, {Fouqu{\'e}}, {Frith}, {Galand}, {Gaulme},
  {Hern{\'a}ndez}, {Grasset}, {Grassi}, {Grenfell}, {Griffin}, {Griffith},
  {Gr{\"o}zinger}, {Guedel}, {Guio}, {Hainaut}, {Hargreaves}, {Hauschildt},
  {Heng}, {Heyrovsky}, {Hueso}, {Irwin}, {Kaltenegger}, {Kervella}, {Kipping},
  {Koskinen}, {Kov{\'a}cs}, {La Barbera}, {Lammer}, {Lellouch}, {Leto}, {Lopez
  Morales}, {Lopez Valverde}, {Lopez-Puertas}, {Lovis}, {Maggio}, {Maillard},
  {Maldonado Prado}, {Marquette}, {Martin-Torres}, {Maxted}, {Miller},
  {Molinari}, {Montes}, {Moro-Martin}, {Moses}, {Mousis}, {Nguyen Tuong},
  {Nelson}, {Orton}, {Pantin}, {Pascale}, {Pezzuto}, {Pinfield}, {Poretti},
  {Prinja}, {Prisinzano}, {Rees}, {Reiners}, {Samuel}, {S{\'a}nchez-Lavega},
  {Forcada}, {Sasselov}, {Savini}, {Sicardy}, {Smith}, {Stixrude},
  {Strazzulla}, {Tennyson}, {Tessenyi}, {Vasisht}, {Vinatier}, {Viti},
  {Waldmann}, {White}, {Widemann}, {Wordsworth}, {Yelle}, {Yung}, \&
  {Yurchenko}}]{Tinetti2012}
{Tinetti}, G., {Beaulieu}, J.~P., {Henning}, T., {et~al.} 2012, Experimental
  Astronomy, 34, 311

\bibitem[{{Vidal-Madjar} {et~al.}(2003){Vidal-Madjar}, {Lecavelier des Etangs},
  {D{\'e}sert}, {Ballester}, {Ferlet}, {H{\'e}brard}, \&
  {Mayor}}]{Vidal-Madjar2003}
{Vidal-Madjar}, A., {Lecavelier des Etangs}, A., {D{\'e}sert}, J.-M., {et~al.}
  2003, \nat, 422, 143

\bibitem[{{Vidal-Madjar} {et~al.}(2011){Vidal-Madjar}, {Sing}, {Lecavelier Des
  Etangs}, {Ferlet}, {D{\'e}sert}, {H{\'e}brard}, {Boisse}, {Ehrenreich}, \&
  {Moutou}}]{Vidal-Madjar2011}
{Vidal-Madjar}, A., {Sing}, D.~K., {Lecavelier Des Etangs}, A., {et~al.} 2011,
  \aap, 527, A110

\bibitem[{{Wu} \& {Lithwick}(2013)}]{Wu2013}
{Wu}, Y. \& {Lithwick}, Y. 2013, \apj, 763, 13

\bibitem[{{Zahnle} {et~al.}(2009){Zahnle}, {Marley}, {Freedman}, {Lodders}, \&
  {Fortney}}]{Zahnle2009}
{Zahnle}, K., {Marley}, M.~S., {Freedman}, R.~S., {Lodders}, K., \& {Fortney},
  J.~J. 2009, \apjl, 701, L20

\end{thebibliography}
\end{multicols}

\end{document}